\documentclass[epj]{webofc}
\usepackage[utf8]{inputenc}
\usepackage[varg]{txfonts}   % Web of Conferences font
    
\newcommand{\cN}{{\cal N}}
\newcommand{\cF}{{\cal F}}
\newcommand{\cV}{{\cal V}}
\newcommand{\cW}{{\cal W}}
\newcommand{\cY}{{\cal Y}}
\newcommand{\cT}{{\cal T}}

\newcommand{\lambdab}{{\overline{\lambda}}}
\newcommand{\phib}{{\overline{\phi}}}

\newcommand{\Tr}{{\rm Tr\;}}
\newcommand{\hf}{\frac{1}{2}}
\newcommand{\qtr}{\frac{1}{4}}
\newcommand{\rtwo}{\sqrt{2}}
\newcommand{\bn}{{\bf n}}

\newcommand{\tQ}{\widetilde{Q}}

\def\nn{\nonumber}
\def\bec{\begin{center}}
\def\eec{\end{center}}
\def\beq{\begin{equation}}
\def\eeq{\end{equation}}
\def\bea{\begin{eqnarray}}
\def\eea{\end{eqnarray}}

\usepackage{subfigure}

\wocname{EPJ Web of Conferences}
\woctitle{Lattice2017}

\begin{document}
%%%%%%%%%%%%%%%%%%%%%%%%%%%%%%%%%%%%%%%%%%%%%%%%%%%%%%%%%%%%%%%%%%%%%%%%%%%%%

\selectlanguage{english}
%%%%%%%%%%%%%%%%%%%%%%%%%%%%%%%%%%%%%%%%%%%%%%%%%%%%%%%%%%%%%%%%%%%%%%%%%%%%%
\title{
$\cN = 2^*$ Yang-Mills on the Lattice
}
%%%%%%%%%%%%%%%%%%%%%%%%%%%%%%%%%%%%%%%%%%%%%%%%%%%%%%%%%%%%%%%%%%%%%%%%%%%%%
\author{
\firstname{Anosh} \lastname{Joseph}\inst{1}\fnsep\thanks{Speaker, \email{anosh.joseph@icts.res.in}}
}
%%%%%%%%%%%%%%%%%%%%%%%%%%%%%%%%%%%%%%%%%%%%%%%%%%%%%%%%%%%%%%%%%%%%%%%%%%%%%
\institute{
International Centre for Theoretical Sciences (ICTS-TIFR), \\
Tata Institute of Fundamental Research, Shivakote, Bangalore 560089 INDIA
}
%%%%%%%%%%%%%%%%%%%%%%%%%%%%%%%%%%%%%%%%%%%%%%%%%%%%%%%%%%%%%%%%%%%%%%%%%%%%%
\abstract{
The $\cN = 2^*$ Yang-Mills theory in four dimensions is a non-conformal theory that appears as a mass deformation of maximally supersymmetric $\cN = 4$ Yang-Mills theory. This theory also takes part in the AdS/CFT correspondence and its gravity dual is type IIB supergravity on the Pilch-Warner background. The finite temperature properties of this theory have been studied recently in the literature. It has been argued that at large $N$ and strong coupling this theory exhibits no thermal phase transition at any non-zero temperature. The low temperature $\cN = 2^*$ plasma can be compared to the QCD plasma. We provide a lattice construction of $\cN = 2^*$ Yang-Mills on a hypercubic lattice starting from the $\cN = 4$ gauge theory. The lattice construction is local, gauge-invariant, free from fermion doubling problem and preserves a part of the supersymmetry. This nonperturbative formulation of the theory can be used to provide a highly nontrivial check of the AdS/CFT correspondence in a non-conformal theory.
}
%%%%%%%%%%%%%%%%%%%%%%%%%%%%%%%%%%%%%%%%%%%%%%%%%%%%%%%%%%%%%%%%%%%%%%%%%%%%%
\maketitle
%%%%%%%%%%%%%%%%%%%%%%%%%%%%%%%%%%%%%%%%%%%%%%%%%%%%%%%%%%%%%%%%%%%%%%%%%%%%%

%%%%%%%%%%%%%%%%%%%%%%%%%%%%%%%%%%%%%%%%%%%%%%%%%%%%%%%
\section{Introduction}
\label{sec:intro}
%%%%%%%%%%%%%%%%%%%%%%%%%%%%%%%%%%%%%%%%%%%%%%%%%%%%%%%

Supersymmetric quantum field theories form an interesting class of theories by themselves. We can construct many phenomenologically relevant models using these theories as starting points. Many of the interesting features of supersymmetric field theories occur when the coupling parameter is of order one. It is difficult to study analytically the strong coupling regimes of quantum field theories in general. If we could formulate them on a spacetime lattice, in a consistent manner, we would have a first principles definition of the theory that can be used to study their nonperturbative regimes. Certain classes of supersymmetric field theories can be formulated on a spacetime lattice by preserving a subset of supersymmetries. These approaches are based on the methods of topological twisting and orbifolding and they can be used to study theories with extended supersymmetries. 

Supersymmetric lattices have been constructed for several classes of theories in various spacetime dimensions \cite{Sugino:2003yb, Catterall:2009it, Catterall:2011pd, Joseph:2011xy, Catterall:2011aa, Catterall:2013roa, Joseph:2013bra, Joseph:2014bwa, Catterall:2014vka, Joseph:2014wxa, Joseph:2015xwa, Joseph:2016tlc, Joseph:2016cdq}, including the well known $\cN = 4$ supersymmetric Yang-Mills theory in four spacetime dimensions.

Here we detail a lattice construction of a very interesting theory, known as $\cN = 2^*$ supersymmetric Yang-Mills (SYM) theory. (See also Ref. \cite{Joseph:2017nap}.) This four-dimensional non-conformal theory is a one-parameter mass deformation of four-dimensional $\cN = 4$ SYM theory. This theory also takes part in the AdS/CFT correspondence. Its gravitational dual has been constructed by Pilch and Warner \cite{Pilch:2000ue}.

In the recent past, supersymmetric lattice constructions have been used to test and validate the gauge-gravity duality conjecture \cite{Catterall:2009xn, Catterall:2010fx, Catterall:2010ya, Kadoh:2015mka, Filev:2015hia, Asano:2016xsf, Berkowitz:2016jlq, Asano:2016kxo, Catterall:2017lub}. See Refs. \cite{Catterall:2011cea, Schaich:2014pda} for computer codes developed for simulating such supersymmetric theories.     

We use the method of topological twisting to construct $\cN = 2^*$ SYM on a Euclidean spacetime lattice. The lattice construction preserves one supersymmetry charge at finite lattice spacing. The lattice formulation is also local, gauge invariant and free from the problem of fermion doublers.      

One could use this construction to explore the nonperturbative regimes of the theory, including its thermodynamic properties, and compare with the results obtained from the gravitational side.  

%%%%%%%%%%%%%%%%%%%%%%%%%%%%%%%%%%%%%%%%%%%%%%%%%%%%%%%
\section{Four-dimensional $\cN = 4$ Yang-Mills}
\label{sec:4d_n4_sym}
%%%%%%%%%%%%%%%%%%%%%%%%%%%%%%%%%%%%%%%%%%%%%%%%%%%%%%%

We consider $\cN = 4$ supersymmetric Yang-Mills theory (SYM) on flat ${\mathbb R}^4$. In the language of $\cN = 1$ superfields $\cN = 4$ SYM theory contains one vector multiplet $V$ and three adjoint chiral multiplets $\Phi_s$, with $s = 1, 2, 3$.

The physical component fields of the superfields are
\beq
\begin{aligned}
V &\longrightarrow A_\mu,~\lambda_{4\alpha},~\lambdab^4_{\phantom{4}\dot{\alpha}};~~ \Phi_s, \Phi^{\dagger s} &\longrightarrow \phi_s,~\lambda_{s\alpha},~\phi^{\dagger s},~\lambdab^s_{\phantom{s}\dot{\alpha}}.
\end{aligned}
\eeq

The global symmetry group of the theory is $SU(2)_L \times SU(2)_R \times SU(4)$, where $SU(2)_L \times SU(2)_R \simeq SO(4)$ is the Euclidean Lorentz rotation group and $SU(4) \simeq SO(6)$ denotes the R-symmetry group. 

The gauge fields $A_\mu$ are scalars under $SU(4)$. The gauginos, $\lambda_{s\alpha}$ and $\lambdab^s_{\phantom{s}\dot{\alpha}}$, and the six scalars, $\phi_s$ and $\phi^{\dagger s}$ transform as ${\mathbf 4} \oplus \overline{{\mathbf 4}}$ and ${\mathbf 6}$, respectively under $SU(4)$ internal rotation symmetry. The scalars can be packaged into an antisymmetric and self-conjugate tensor $\phi_{uv}$ ($u, v = 1, 2, 3,4$) representing the indices of the fundamental representation of $SU(4)$. In this notation the gauginos of vector and chiral multiplets can be combined: $\lambda_{u\alpha},~\lambdab^u_{\phantom{u}\dot{\alpha}}$. All fields are in the adjoint representation of the gauge group $G$. Here we take $G$ to be $SU(N)$.

%%%%%%%%%%%%%%%%%%%%%%%%%%%%%%%%%%%%%%%%%%%%%%%%%%%%%%%%%%%%%
\section{Mass Deformation and $\cN = 2^*$ Yang-Mills}
\label{sec:mass_deform}
%%%%%%%%%%%%%%%%%%%%%%%%%%%%%%%%%%%%%%%%%%%%%%%%%%%%%%%%%%%%%

We can combine the superfield $V$ and one of the adjoint chiral superfields say, $\Phi_3$ to form an $\cN = 2$ vector multiplet. We can combine the chiral superfields $\Phi_1$ and $\Phi_2$ to form an $\cN = 2$ hypermultiplet. The $\cN = 2^*$ SYM theory is a one-parameter (real) mass deformation of $\cN = 4$ SYM obtained by giving mass to the fields of $\cN = 2$ hypermultiplet. The mass terms break supersymmetry softly from $\cN = 4$ to $\cN = 2$. The $\cN = 2^*$ SYM theory has a fixed point in the far UV, which is the conformal $\cN = 4$ SYM theory. The mass deformation is relevant and it induces running in the coupling, so that in the deep IR the theory becomes pure $\cN = 2$ SYM theory.

The mass deformation takes the following form, on flat ${\mathbb R}^4$, in terms of the component fields \cite{Labastida:1997xk, Buchel:2012gw}
\bea
\label{eq:n2-star-untwisted-mass}
S_m &=& \frac{1}{g^2} \int d^4x~ \Tr  \Big( - m \lambda_1^{\phantom{1}\alpha} \lambda_{2\alpha} - m \lambdab^1_{\phantom{1}\dot{\alpha}} \lambdab^{2\dot{\alpha}} + m^2 \phi_1^{\phantom{\dagger}} \phi_1^\dagger + m^2 \phi_2^{\phantom{\dagger}} \phi_2^\dagger \nn \\
&& \quad \quad - \rtwo m \phi_3^{\phantom{\dagger}}[\phi_1^{\phantom{\dagger}}, \phi_1^\dagger] - \rtwo m \phi_3 [\phi_2^{\phantom{\dagger}}, \phi_2^\dagger] - \rtwo m \phi_3^\dagger[\phi_1^{\phantom{\dagger}}, \phi_1^\dagger] - \rtwo m \phi_3^\dagger [\phi_2^{\phantom{\dagger}}, \phi_2^\dagger] \Big),
\eea
where $m$ is the mass parameter and $g$ the coupling constant of the theory. The mass deformation gives conventional mass terms for 2 Weyl fermions and 2 complex scalars and also tri-linear couplings between the $\cN = 2$ hypermultiplet scalars and the vector multiplet scalar $\phi_3$. 

The action of the $\cN = 2^*$ SYM can be expressed as \cite{Hoyos:2011uh, Buchel:2012gw}
\beq
S_{\cN=2^*} = S_{\cN=4} + S_m,
\eeq
where $S_{\cN=4}$ is the $\cN=4$ SYM action

The theory has an $SU(2) \times U(1)$ R-symmetry. The symmetry breaking gives equal masses to 2 of the 4 Weyl fermions. The $SU(2)$ acts on the 2 massless fermions, and the $U(1) \simeq SO(2)$ mixes the 2 massive fermions. 

%%%%%%%%%%%%%%%%%%%%%%%%%%%%%%%%%%%%%%%%%%%%%%%%%%%%
\section{Vafa-Witten Twist of $\cN = 4$ Yang-Mills}
\label{sec:Vafa-Witten-twist}
%%%%%%%%%%%%%%%%%%%%%%%%%%%%%%%%%%%%%%%%%%%%%%%%%%%%

We are interested in formulating $\cN = 2^*$ Yang-Mills theory on a Euclidean spacetime lattice. We begin with $\cN = 4$ Yang-Mills theory on flat ${\mathbb R}^4$. We are interested in discretizing a twisted version of $\cN = 2^*$ Yang-Mills theory. The process of twisting produces at least one nilpotent supersymmetry and this part of the supersymmetry algebra can be easily transported onto the lattice by preserving part of the supersymmetries. We are interested in the Vafa-Witten twist of the $\cN = 4$ Yang-Mills theory \cite{Vafa:1994tf}. For this particular twist, the internal symmetry group $SU(4)$ is decomposed as $SU(2)_F \times SU(2)_I$ such that the twisted global symmetry group is 
\beq
SU(2)'_L \times SU(2)_R \times SU(2)_F,
\eeq
where
\beq
SU(2)'_L = {\rm diag}~\Big(SU(2)_L \times SU(2)_I\Big),
\eeq
and $SU(2)_F$ remains as a residual internal symmetry group. The fields and supercharges of the untwisted theory are rewritten in terms of the twisted fields. 

The process of twisting gives rise to the following fields. In the bosonic sector we have a gauge field $A_\mu$, a self-dual 2-form $B_{\mu\nu}$ and three scalars $\phi, \phib, C$. The fermions are two 0-forms $\eta, \zeta$ two 1-forms $\psi_\mu, \chi_\mu$ and two 2-forms $\chi_{\mu\nu}, \psi_{\mu\nu}$. We note that when the theory is formulated on a flat manifold or in general on a hyper-Kahler manifold, the twisted theories coincide with the physical theory \cite{Vafa:1994tf}.

The subset of the twisted fields $(A_\mu, \phi, \phib, \eta, \psi_\mu, \chi_{\mu\nu})$ can be readily recognized as the twisted vector multiplet of the four-dimensional $\cN = 2$ SYM (Donaldson-Witten theory) \cite{Witten:1988ze}. The twisted theory contains an $\cN = 2$ hypermultiplet with the field content $(C, B_{\mu\nu}, \zeta, \chi_\mu, \psi_{\mu\nu})$. We make this hypermultiplet massive when we construct the twisted $\cN = 2^*$ SYM theory.

%%%%%%%%%%%%%%%%%%%%%%%%%%%%%%%%%%%%%%%%%%%%%%%%%%%%%%%%
\subsection{$Q$ Supersymmetry Transformations}
\label{subsec:Q-susy-trafos}
%%%%%%%%%%%%%%%%%%%%%%%%%%%%%%%%%%%%%%%%%%%%%%%%%%%%%%%%

The twisting procedure gives rise to the following twisted supercharges: two scalars ($Q, \tQ$), two vectors ($Q_\mu, \tQ_\mu$) and two self-dual tensors ($Q_{\mu\nu}, \tQ_{\mu\nu}$). All twisted supercharges leave the twisted $\cN = 4$ SYM action invariant. 

We are interested in the scalar supercharges $Q$ and $\tQ$. Introducing two auxiliary fields, $H_\mu$ and $H_{\mu\nu}$, the off-shell action of the $Q$ supercharge on the twisted fields takes the following form
\begin{align}
 Q A_\mu &= - \psi_\mu, & Q C &= \rtwo \zeta, \nn \\
 Q \psi_\mu &= - 2 \rtwo D_\mu \phi, & Q \zeta &= - 2 [\phi, C], \nn \\
 Q \phi &= 0,   & & \nn \\
 Q \phib &= \rtwo \eta,   & Q \chi_\mu &= 2 H_\mu, \\
 Q \eta &= - 2 [\phi, \phib],   & Q H_\mu &= - \rtwo [\phi, \chi_\mu], \nn \\
 Q \chi_{\mu\nu} &= 2 H_{\mu\nu},   & Q B_{\mu\nu} &= \rtwo \psi_{\mu\nu}, \nn \\
 Q H_{\mu\nu} &= - \rtwo [\phi, \chi_{\mu\nu}] &  Q \psi_{\mu\nu} &= - 2 [\phi, B_{\mu\nu}]. \nn
\end{align}

The $Q$ supercharge satisfies the algebra $Q^2 = \delta^g_\gamma$ with $\delta^g_\gamma$ denoting the gauge transformation with parameter $\gamma = 2 \rtwo \phi$. Similarly we can write down the $\tQ$ supersymmetry transformations of the twisted theory. We have $\tQ^2 = \delta^g_{-\gamma}$.

%%%%%%%%%%%%%%%%%%%%%%%%%%%%%%%%%%%%%%%%%%%%%%
\subsection{Twisted Action}
\label{subsec:twisted-act}
%%%%%%%%%%%%%%%%%%%%%%%%%%%%%%%%%%%%%%%%%%%%%%

We can obtain the twisted action of the $\cN = 4$ SYM through successive variations of $Q$ and $\tQ$ on a functional $\cF$ known as the action potential \cite{Labastida:1997vq, Labastida:1997xk}. We have the twisted action
\beq
S_{\cN=4} = \frac{1}{g^2} \int d^4x~ Q \tQ ~\cF,
\eeq
where the action potential
\bea
\cF &=& \Tr \Big( -\frac{1}{2 \rtwo} B_{\mu\nu} F_{\mu\nu} - \frac{1}{24 \rtwo} B_{\mu\nu} [B_{\mu\rho}, B_{\nu\rho}] - \frac{1}{8} \chi_{\mu\nu} \psi_{\mu\nu} - \frac{1}{8} \psi_\mu \chi_\mu - \frac{1}{8} \eta \zeta \Big).   
\eea

The Vafa-Witten twisted action can be written as the $Q$ variation of a gauge fermion $\Psi$ (which in turn is the $\tQ$ variation of $\cF$)
\beq
S_{\cN=4} = \frac{1}{g^2} \int d^4x~ Q \Psi,
\eeq
with $\Psi$ taking the form
\bea
\Psi &=& \Tr \Big( \chi_{\mu\nu} \Big[\hf F_{\mu\nu} + \qtr H_{\mu\nu} + \frac{1}{8} [B_{\mu\rho}, B_{\nu\rho}] + \qtr [C, B_{\mu\nu}] \Big] + \frac{1}{2\rtwo} \psi_\mu (D_\mu \phib) - \qtr \eta [\phi, \phib] \nn \\
&& \quad \quad - \qtr \zeta [C, \phib] - \frac{1}{4} \psi_{\mu\nu} [B_{\mu\nu}, \phib] + \chi_\mu \Big[ \qtr H_\mu - \frac{1}{2\rtwo} (D_\mu C) - \frac{1}{2 \rtwo} (D_\nu B_{\nu\mu}) \Big] \Big).   
\eea

Applying $Q$ variation on the gauge fermion we obtain the twisted $\cN = 4$ SYM action.

%%%%%%%%%%%%%%%%%%%%%%%%%%%%%%%%%%%%%%%%%%%%%%%%%%%%%%%%%%%
\section{Twisted $\cN = 2^*$ Yang-Mills}
\label{sec:twisted-n2*-sym}
%%%%%%%%%%%%%%%%%%%%%%%%%%%%%%%%%%%%%%%%%%%%%%%%%%%%%%%%%%%

Once we know the transformations from the untwisted fields to twisted fields it is easy to write down the action of the $\cN = 2^*$ SYM theory in the twisted language. The $\cN = 2^*$ SYM theory is obtained by giving masses to the $\cN = 2$ hypermultiplet fields $(C, B_{\mu\nu}, \zeta, \chi_\mu, \psi_{\mu\nu})$.

We can rewrite the mass terms of the untwisted theory using twisted variables. We have
\beq
S_{\cN=2^*} = S_{\cN=4} + S_m,
\eeq
where $S_{\cN=4}$ is the twisted $\cN=4$ SYM action and $S_m$ has the form
\bea
S_m &=& \frac{1}{g^2} \int d^4x~ \Tr \Big[ - \hf m^2 B_{\mu\nu}^2 - \hf m^2 C^2 + \frac{im}{\rtwo} (\psi_{12}\psi_{23} + \psi_{13} \zeta) - \frac{im}{\rtwo} (\chi_1 \chi_2 - \chi_3 \chi_4)\nn \\
&& \quad \quad - \hf m \phi \Big([B_{\mu\nu}, B_{\mu\nu}] + [C, C] \Big) - \hf m \phib \Big([B_{\mu\nu}, B_{\mu\nu}] + [C, C] \Big) \nn \\
&& \quad \quad + i m \phi \Big([B_{12}, B_{23}] + [B_{13}, C]\Big) + i m \phib \Big([B_{12}, B_{23}] + [B_{13}, C]\Big) \Big].
\eea

From the above form of the twisted $\cN = 2^*$ SYM action we note the following: $(i.)$ The net $U(1)_R$ charge of $S_m$ is non-zero. There are mass terms with $U(1)_R$ charge $-2$, $0$ and $+2$, $(ii.)$ There are mass terms that are not invariant under twisted Lorentz symmetry, $(iii.)$ The piece $S_m$ breaks the exchange symmetry under $Q \leftrightarrow \tQ$. This is expected since we have given mass to only one of the $\cN = 2$ hypermutiplets.

%%%%%%%%%%%%%%%%%%%%%%%%%%%%%%%%%%%%%%%%%%%%%%%%%%%%%%%%%%%
\subsection{Mass-dependent $\tQ$ and $Q$ Transformations}
\label{subsec:mass-dep-q-qbar}
%%%%%%%%%%%%%%%%%%%%%%%%%%%%%%%%%%%%%%%%%%%%%%%%%%%%%%%%%%%

We would like to write down the $\cN = 2^*$ SYM action in a $Q$-exact form, with an appropriate gauge fermion. In order to achieve this we need to modify the $Q$ and $\tQ$ transformations on the twisted fields in a mass dependent way.

We modify the off-shell $Q$ transformations in a mass dependent way. Defining the modified supercharge $Q^{(m)}$ we have the following transformations

\begin{align}
Q^{(m)} A_\mu &= - \psi_\mu, & Q^{(m)} \psi_\mu &= -2 \rtwo D_\mu \phi, \nn \\
Q^{(m)} \phi &= 0, & & \nn \\
Q^{(m)} \phib &= \rtwo \eta, & Q^{(m)} \eta &= - 2 [\phi, \phib], \nn \\
Q^{(m)} C &= \rtwo \zeta, & Q^{(m)} \zeta &= - 2 [\phi, C] + 2 m C, \\
Q^{(m)} \chi_\mu &= 2 H_\mu, & Q^{(m)} H_\mu &= - \rtwo [\phi, \chi_\mu] + \rtwo m \chi_\mu, \nn \\
Q^{(m)} B_{\mu\nu} &= \rtwo \psi_{\mu\nu}, & Q^{(m)} \psi_{\mu\nu} &= - 2 [\phi, B_{\mu\nu}] + 2 m B_{\mu\nu}, \nn \\
Q^{(m)} \chi_{\mu\nu} &= 2 H_{\mu\nu}, & Q^{(m)} H_{\mu\nu} &= - \rtwo [\phi, \chi_{\mu\nu}]. \nn 
\end{align}

The $Q^{(m)}$ supercharge satisfies the algebra: $(Q^{(m)})^2 X = 2 \rtwo [X, \phi] + 2 \rtwo m \alpha X$, with $\alpha = 1$ for the fields $X = \{ \zeta, \chi_\mu, \psi_{\mu\nu}, C, H_\mu, B_{\mu\nu} \}$ and $\alpha = 0$ for the rest of the fields.

It would be interesting to see if the deformation part of the algebra represents rotation by an R-symmetry generator. Similar topics were considered in Ref. \cite{Hanada:2010kt} by Hanada, Matsuura and Sugino. 

We can obtain the $\cN = 2^*$ SYM action as a $Q^{(m)}$ variation of the following modified gauge fermion
\bea
\Psi^{(m)} &=& \Tr \Big(\chi_{\mu\nu} \Big[\hf F_{\mu\nu} - \qtr H_{\mu\nu} - \frac{1}{8} [B_{\mu\rho}, B_{\nu\rho}] - \frac{1}{4} [C, B_{\mu\nu}] \Big] + \frac{1}{2\rtwo} \psi_\mu (D_\mu \phib) - \frac{1}{4} \eta [\phi, \phib] \nn \\
&& \quad \quad + (\cV + \cW + \cY) - \qtr \zeta [C, \phib] - \qtr \psi_{\mu\nu} [B_{\mu\nu}, \phib] + \cT \nn \\
&&\quad \quad + \chi_\mu \Big[- \frac{1}{2 \rtwo} (D_\mu C) - \frac{1}{2\rtwo} (D_\nu B_{\nu\mu}) \Big] \Big),
\eea
where $\cV \equiv -\qtr m \Big( (\psi_{12} - i \psi_{23}) (B_{12} + i B_{23}) + (\psi_{13} - i \zeta)(B_{13} + i C)\Big), \cW \equiv \frac{i}{4} \Big( - \psi_{12} [\phib, B_{23}] + \psi_{23} [\phib, B_{12}] + \eta [B_{12}, B_{23}] \Big), \cY \equiv \frac{i}{4} \Big( - \psi_{13} [\phib, C] + \zeta [\phib, B_{13}] + \eta [B_{13}, C] \Big)$ and $\cT \equiv \qtr \Big( (\chi_1 - i \chi_2)(H_1 + iH_2) + (\chi_3 + i \chi_4)(H_3 - iH_4) \Big)$.

It is straightforward to show that the $Q^{(m)}$ variation of $\Psi^{(m)}$ will produce the twisted action of $\cN = 2^*$ SYM 
\beq
S_{\cN=2^*} = \frac{1}{g^2} \int d^4x~ Q^{(m)}\Psi^{(m)}.
\eeq

%%%%%%%%%%%%%%%%%%%%%%%%%%%%%%%%%%%%%%%%%%%%%%%%%%%
\section{Lattice Formulation}
\label{sec:lattice}
%%%%%%%%%%%%%%%%%%%%%%%%%%%%%%%%%%%%%%%%%%%%%%%%%%%

%%%%%%%%%%%%%%%%%%%%%%%%%%%%%%%%%%%%%%%%%%%%%%%%%%%
\subsection{Balanced Topological Field Theory Form}
%%%%%%%%%%%%%%%%%%%%%%%%%%%%%%%%%%%%%%%%%%%%%%%%%%%

We can rewrite the Vafa-Witten twisted $\cN = 4$ SYM theory in a form known as the balanced topological field theory (BTFT) form. The existence of two scalar supercharges $Q$ and $\tQ$ would allow us to express the $\cN = 4$ theory in this form. In Ref. \cite{Dijkgraaf:1996tz} Dijkgraf and Moore wrote down the BTFT form of the Vafa-Witten twisted theory. Sugino has used this approach to formulate four-dimensional $\cN = 4$ and $\cN = 2$ SYM theories on a hypercubic lattice \cite{Sugino:2003yb}.

We can define a three component vector $\vec{\Phi}$, which is a function of the field strength. The components of this vector take the form $\Phi_A \equiv 2 ( F_{A4} + 1/2 \epsilon_{ABC} F_{BC})$, with $A, B, C = 1, 2, 3$. Similarly we introduce 3-component vector fields $\vec{B}$, $\vec{H}$, $\vec{\psi}$ and $\vec{\chi}$. 

The action potential takes the following form using the BTFT notation
\beq
\cF = \Big( - \frac{1}{2 \rtwo} B_A \Phi_A - \frac{1}{24 \rtwo} \epsilon_{ABC} B_A [B_B, B_C] - \frac{1}{8} \chi_A \psi_A - \frac{1}{8} \psi_\mu \chi_\mu - \frac{1}{8} \eta \zeta \Big),
\eeq
and the twisted action can be expressed in BTFT form, arising from $\tQ$ and $Q$ variations of $\cF$.

%%%%%%%%%%%%%%%%%%%%%%%%%%%%%%%%%%%%%%%%%%%%%%%%%%%
\subsection{Lattice Regularized Theory}
%%%%%%%%%%%%%%%%%%%%%%%%%%%%%%%%%%%%%%%%%%%%%%%%%%%

We formulate the theory on a four-dimensional hypercubic lattice following the discretization prescription given by Sugino \cite{Sugino:2003yb}. In the lattice theory the gauge fields $A_\mu$ are promoted to compact unitary variables
\beq
U_\mu(\bn) \equiv U(\bn, \bn + \mu) = e^{A_\mu(\bn)},~~U^\dagger_\mu(\bn - \mu) \equiv U(\bn, \bn - \mu) = e^{-A_\mu(\bn)}
\eeq
on the link $(\bn, \bn+\mu)$. All other variables are distributed on the sites. 

Upon using the language of BTFT form we have the $Q^{(m)}$ transformations on the lattice
\begin{align}
Q^{(m)} U_\mu(\bn) &= - \psi_\mu U_\mu(\bn),   & Q^{(m)} \psi_\mu (\bn) &= \psi_\mu(\bn) \psi_\mu(\bn) - 2 \rtwo D_\mu^{(+)} \phi(\bn), \nn \\
Q^{(m)} \phi(\bn)  &= 0,          & \nn \\
Q^{(m)} \phib(\bn) &= \rtwo \eta(\bn), & Q^{(m)} \eta(\bn) &= - 2 [\phi(\bn), \phib(\bn)], \nn \\
Q^{(m)} C(\bn) &= \rtwo \zeta(\bn), & Q^{(m)} \zeta(\bn) &= - 2 [\phi(\bn), C(\bn)] + 2 m C(\bn), \\
Q^{(m)} \chi_\mu(\bn) &= 2 H_\mu(\bn), & Q^{(m)} H_\mu(\bn) &= - \rtwo [\phi(\bn), \chi_\mu(\bn)] + \rtwo m \chi_\mu(\bn), \nn \\
Q^{(m)} B_A(\bn) &= \rtwo \psi_A(\bn), & Q^{(m)} \psi_A(\bn) &= - 2 [\phi(\bn), B_A(\bn)] + 2 m B_A(\bn), \nn \\
Q^{(m)} \chi_A(\bn) &= 2 H_A(\bn), & Q^{(m)} H_A(\bn) &= - \rtwo [\phi(\bn), \chi_A(\bn)]. \nn
\end{align}

These transformations were originally proposed by Sugino in Ref. \cite{Sugino:2003yb}, for the $m = 0$ case, while formulating the $\cN = 4$ and $\cN = 2$ SYM theories on the lattice. 

In the above transformations $D^{(+)}$ and $D^{(-)}$ are the forward and backward covariant difference operators, respectively: $D_\mu^{(+)} f(\bn) = U_\mu(\bn) f(\bn + \mu) U^\dagger_\mu(\bn) - f(\bn),~~ D^{(-)}_\mu g_\mu(\bn) = g_\mu(\bn) - U^\dagger_\mu(\bn - \mu) g_\mu(\bn - \mu) U_\mu(\bn - \mu)$.

The $Q^{(m)}$ transformations reduce to their continuum counterparts in the limit of vanishing lattice spacing. The term quadratic in $\psi_\mu$ is suppressed by additional power of the lattice spacing. $\left(Q^{(m)}\right)^2$ on the lattice obeys a relation similar to the one given in the continuum.

Once we have the $Q^{(m)}$ transformation rule closed among lattice variables, it is almost straightforward to construct the lattice action. 

The functional $\Phi_A$ takes the following form on the lattice \cite{Sugino:2003yb}
\beq
\Phi_A(\bn) = - \Big(U_{A4}(\bn) - U_{4A}(\bn) + \hf \sum_{B, C = 1}^3 \epsilon_{ABC} (U_{BC}(\bn) - U_{CB}(\bn))\Big).
\eeq

The plaquette variables $U_{\mu\nu}(x)$ are defined as: $U_{\mu\nu}(\bn) \equiv U_\mu(\bn) U_\nu(\bn + \mu) U_\mu(\bn + \nu)^\dagger U_\nu(\bn)^\dagger$.

We can integrate out the auxiliary field $\vec{H}(\bn)$ so that the $\vec{\Phi}(\bn)^2$ term gives the gauge kinetic term on the lattice
\beq
\frac{1}{2 g_0^2} \sum_\bn \sum_{\mu < \nu} \Tr \Big[ -(U_{\mu\nu}(\bn) - U_{\nu\mu}(\bn))^2 \Big].
\eeq

We note that the above term contains double winding plaquette terms. On the other hand, the standard Wilson action has a unique minimum $U_{\mu\nu} = 1$. 

The lattice action has many classical vacua: the center elements of $SU(N)$ and also the configurations $U_{\mu\nu} = {\rm diag} (\pm 1, \cdots, \pm 1)$, up to gauge transformations, with any combinations of $\pm 1$, with `$-1$' appearing even times are allowed in the diagonal entries. 

It has some serious consequences. Since the diagonal entries can be taken freely for each plaquette, it results in a huge degeneracy of vacua with the number growing as exponential of the number of plaquettes. We need to add up contributions from all the minima in order to see the dynamics of the model. In this case, the ordinary weak field expansion around a single vacuum $U_{\mu\nu} = 1$ cannot be justified. That is, we are unable to say anything about the continuum limit of the lattice theory without its nonperturbative investigations.  

We could add a term proportional to the standard Wilson action to the lattice action in order to resolve the degeneracy
\beq
\Delta S = \frac{\rho}{2 g_0^2} \sum_\bn \sum_{\mu < \nu} \Tr \Big[2 - U_{\mu\nu}(\bn) - U_{\nu\mu}(\bn)\Big],
\eeq
where $\rho$ is a parameter to be tuned. This term resolves the degeneracy with the split $4 \rho/g_0^2$ \cite{Sugino:2003yb}. However, this term breaks $Q^{(m)}$ supersymmetry, even though it justifies the expansion around the vacuum $U_{\mu\nu} = 1$.

The $\cN = 2^*$ SYM action has the following form on the lattice
\beq
S_{\cN=2^*} = \beta_L \sum_\bn Q^{(m)} \Psi^{(m)}(\bn),
\eeq
with $\beta_L$ denoting the lattice coupling. 

It is possible to show that the lattice theory has no fermion doubling problem by following an analysis similar to the one given in \cite{Sugino:2003yb}. We note that the lattice action of $\cN = 2^*$ SYM formulated here is: $(i.)$ gauge invariant, $(ii.)$ local, $(iii.)$ doubler free and $(iv.)$ exact supersymmetric under one supersymmetry charge.

We also note that it would be possible to impose the admissibility condition \cite{Sugino:2004qd} $|| 1 - U_{\mu\nu} || < \epsilon$, with $\epsilon$ a parameter to be determined, on each plaquette variable in order to solve the issues with vacuum degeneracy while keeping supersymmetry. The admissibility condition is imposed on the gauge fermion and it does not affect the $Q$-exact structure of the theory. Ref. \cite{Matsuura:2014pua} discusses another method to avoid the degeneracy while preserving $Q$ supersymmetry.

%%%%%%%%%%%%%%%%%%%%%%%%%%%%%%%%%%%%%%%%%%%%%%%%%%%%%%%%%%%%%%%%%%%%%%
\section{Conclusions}
%%%%%%%%%%%%%%%%%%%%%%%%%%%%%%%%%%%%%%%%%%%%%%%%%%%%%%%%%%%%%%%%%%%%%%

We have provided a lattice construction of $\cN = 2^*$ SYM that respects gauge invariance, locality, and supersymmetry invariance under one supercharge. The formulation is also free from fermion doubling problem. We have also provided the continuum twisted formulation of $\cN = 2^*$ SYM. The nonperturbative construction of $\cN = 2^*$ SYM discussed here can be used to simulate the theory at any finite value of the gauge coupling, mass parameter and number of colors. It would be interesting to simulate the lattice theory and study the observables related to the AdS/CFT correspondence.    

%%%%%%%%%%%%%%%%%%%%%%%%%%%%%%%%%%%%%%%%%%%%%%%%%%%%%%%%%%%%%%%%%%%%%%
{\it Acknowledgements:} We thank discussions with Poul Damgaard, Victor Giraldo, So Matsuura and especially with Fumihiko Sugino. We gratefully acknowledge support from the International Centre for Theoretical Sciences (ICTS-TIFR), the Infosys Foundation and the Indo-French Centre for the Promotion of Advanced Research (IFCPAR/CEFIPRA).
%%%%%%%%%%%%%%%%%%%%%%%%%%%%%%%%%%%%%%%%%%%%%%%%%%%%%%%%%%%%%%%%%%%%%%

\bibliography{Lattice2017_318_Joseph}

\end{document}